\documentclass{article}

\usepackage{graphicx}
\usepackage[margin=1.4in]{geometry}
\usepackage{amsfonts}
\usepackage{amsmath}
\usepackage[english]{babel}
\usepackage{hyperref}
\usepackage{color}
\hypersetup{colorlinks=true,citecolor=blue,linkcolor=blue,filecolor=blue,urlcolor=blue}
\usepackage{tikz}
\usepackage{framed}
\usepackage{setspace}
\usepackage{nicefrac}

\newcommand{\bv}{\begin{verse}}
\newcommand{\ev}{\end{verse}}
\newcommand{\be}{\begin{equation}}
\newcommand{\ee}{\end{equation}}
\newcommand{\bea}{\begin{eqnarray}}
\newcommand{\eea}{\end{eqnarray}}
\newcommand{\bq}{\begin{quotation}}
\newcommand{\eq}{\end{quotation}}

\newcommand*{\Friendone}{\mathsf{\bar F}}
\newcommand*{\Friendtwo}{\mathsf{F}}

\newcommand*{\Assistant}{\mathsf{\bar W}}
\newcommand*{\Wigner}{\mathsf{W}}

\newcommand*{\ExpBegin}[1]{
\vspace{0.2ex}
\begin{center}
\begin{minipage}{\linewidth}
\begin{framed}
\vspace{-0.6ex}
{\centering {\bf #1} \\ }
\vspace{1.1ex}
\nobreak}

\newcommand*{\ExpEnd}{
\vspace{-0.8ex}
\end{framed}
\end{minipage}
\end{center}
}

\begin{document}

%\title{Can Wigner's friend be both a physical system and an agent?}
\title{Respecting One's Fellow: QBism's Analysis of \\ Wigner's Friend\bigskip}
%\large Responses to Baumann and Brukner and Frauchiger and Renner}
%% \title{Can two QBist agents experience one thing? How Wigner and his friend reason about each other}
\author{John B. DeBrota,$^{1,3}$ Christopher A. Fuchs,$^{1,3}$ and R\"udiger Schack$^{2,3}$\medskip
\\
\small $^1$Department of Physics, University of Massachusetts Boston
\\
\small 100 Morrissey Boulevard, Boston MA 02125, \ USA \medskip
\\
\small $^2$Department of Mathematics, Royal Holloway, University of London
\\
\small Egham, Surrey TW20 0EX, United Kingdom \medskip
\\
\small $^3$Stellenbosch Institute for Advanced Study (STIAS)
\\
\small Wallenberg Research Centre, Marais Street, Stellenbosch 7600, South Africa}

\date{5 July 2020}
\maketitle

% \begin{abstract}
\begin{abstract}
According to QBism, quantum states, unitary evolutions, and measurement
operators are all understood as personal judgments of the agent using the
formalism. Meanwhile, quantum measurement outcomes are understood as the
personal experiences of the same agent. Wigner's conundrum of the
friend, in which two agents ostensibly have different accounts of whether or not there is a measurement outcome, thus poses no paradox for QBism.
Indeed the resolution of
Wigner's original thought experiment was central to the development of QBist
thinking. The focus of this paper concerns two very instructive modifications
to Wigner's puzzle: One, a recent no-go theorem by Frauchiger and Renner
\cite{FR}, and the other a thought experiment by Baumann and Brukner
\cite{BB}. We show that the paradoxical features emphasized in these works
disappear once both friend and Wigner are understood as agents on an equal
footing with regard to their individual uses of quantum theory. Wigner's action
on his friend then becomes, from the friend's perspective, an action the friend
takes on Wigner. Our analysis rests on a kind of quantum Copernican principle:
When two agents take actions on each other, each agent has a dual role as a
physical system for the other agent. No user of quantum theory is more
privileged than any other.  In contrast to the sentiment of Wigner's original
paper, neither agent should be considered as in ``suspended animation.''  In
this light, QBism brings an entirely new perspective to understanding Wigner's
friend thought experiments.
\end{abstract}

% \section{Introduction}
\section{Introduction}

Wigner's famous thought experiment \cite{Wigner1961} is a tale of two agents. One agent, Wigner's friend, performs a quantum measurement in a
lab and obtains an outcome. The other agent, Wigner, treats the lab containing his friend and the friend's experimental setup as one large
quantum system and writes down a joint quantum state which evolves continuously in time. Thus for Wigner there is no measurement outcome. Who
is right, Wigner or his friend? There is a difficulty here if one thinks of a measurement outcome as something objective in the sense that it
can be verified in principle by anybody.

The lesson QBism \cite{Fuchs10a,Fuchs13a,Fuchs14a} draws from Wigner's thought experiment is that, for consistency's sake, measurement
outcomes must be regarded as personal to the agent who makes the measurement. This idea first appeared in 2006 in Ref.~\cite{Caves07}, where
the phrase ``facts for the agent'' was coined. Its connection with Wigner's friend was eventually
spelled out fully in 2010 with Ref.~\cite{Fuchs10a}, and the personal nature of an agent's measurement outcomes was further emphasized in
Ref.~\cite{Fuchs14a}, where outcomes were identified with the ``experiences''
of the agent doing the measurement.  As emphasized, e.g., by
Pusey~\cite{Pusey2018}, Wigner's friend thought experiments thus pose no problem
for QBism. In fact, Wigner's friend was central to the development of QBist thinking~\cite{Samizdat2}.

Recently, variations of Wigner's original thought experiment were introduced by Brukner
\cite{Brukner1507,Brukner1804}, Frauchiger and Renner (FR)
\cite{FR}, and Baumann and Brukner (BB) \cite{BB}.\footnote{There are too many
  responses to these papers to cite here, but a sampling of those which attempt to analyse QBism's relation to the thought experiments can be found in Refs.\ \cite{Healey2018,Nurgalieva2018,Proietti2019,Krismer2018,Boge2019,Sudbery2019,Evans2019,Suarez2019,Stacey2019b}. Though Refs.\ \cite{Suarez2019,Stacey2019b} are both very relevant to QBist interests, neither of these get at the heart of the argument made here.} In
Brukner's thought experiment, first described in 2015~\cite{Brukner1507} and
thoroughly analyzed in Ref.~\cite{Brukner1804}, the assumption that measurement
outcomes are objective leads to a Bell inequality and thus to a conflict with
the predictions of quantum mechanics.  Brukner concludes from this that
measurement outcomes should be regarded as ``facts relative to the observer,''
the same conclusion QBism reached by considering the original Wigner's
friend thought experiment.

The main innovation in the very interesting FR scenario is that both Wigner and
a friend make predictions about the outcome of one and the same measurement,
which is performed by Wigner. A seemingly straightforward application of the
quantum formalism then appears to show that the predictions of Wigner and
friend are mutually contradictory. Frauchiger and Renner turn this into a
formal contradiction by postulating a small number of what they believe to be
intuitive assumptions. They conclude that any interpretation of quantum
mechanics has to abandon at least one of these assumptions.

From a QBist perspective, however, there is a fundamental problem with Frauchiger and Renner's analysis.
In their thought
experiment, both Wigner and the friend are agents applying the quantum
formalism, but Frauchiger and Renner treat them in an asymmetric way. Because
Wigner is outside the lab that contains the friend, this asymmetry seems to be
inherent in the very setup of the experiment. We will show that this is not
so. QBism both requires and makes possible a fully symmetric treatment of
Wigner and his friend.  Wigner's action on his friend then becomes, from the
friend's perspective, an action the friend takes on Wigner. Once this is taken
into account, the paradoxical features of the FR thought experiment disappear.

\begin{figure}
    \begin{center}
    \includegraphics[width=\linewidth]{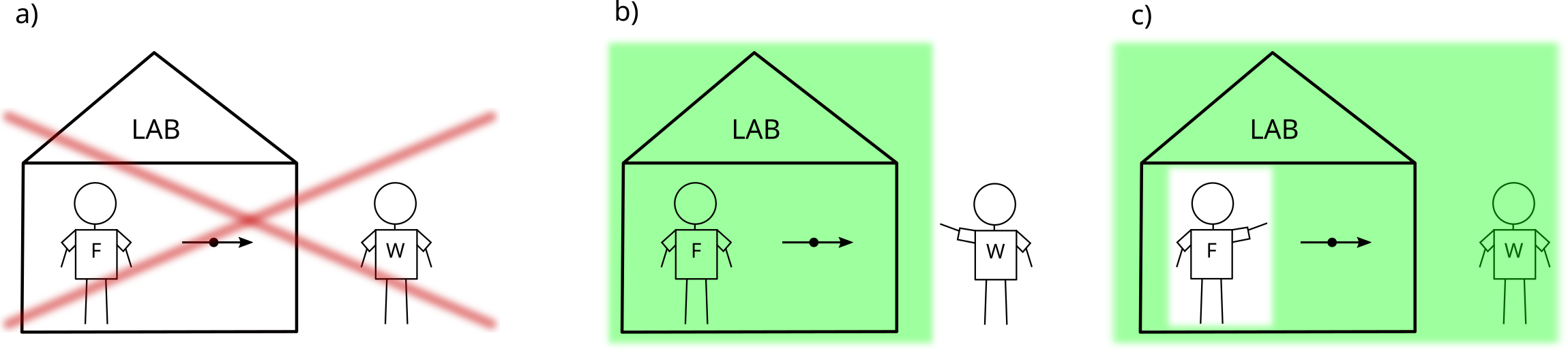}
\end{center}
\caption{\label{ThreeHouses} (a) In usual descriptions of various Wigner's-friend thought experiments, there is an urge to portray everything from a God's eye view. Here, we depict the BB thought experiment, where Wigner, his friend, and a spin-1/2 particle interact, and we symbolize QBism's disapproval of such portrayals with a big red X\@. In QBism, the quantum formalism is only used by agents who stand {\it within\/} the world; there is no God's-eye view. (b) Instead, in QBism, to make predictions, Wigner treats his friend, the particle, and the laboratory surrounding her (all shaded in green) as a physical system external to himself. While (c) to make her own predictions, the friend must reciprocally treat Wigner, the particle, and her surrounding laboratory (all shaded in green) as a physical system external to herself. It matters not that the laboratory spatially surrounds the friend; it, like the rest of the universe, is external to her agency, and that is what counts.}
\label{}
\end{figure}

The more recent BB scenario is similar to the FR thought experiment in that,
again, both Wigner and friend make predictions about the outcome of Wigner's
measurement. Baumann and Brukner appear to show that applying the standard
quantum formalism leads the friend to make a bad prediction. As in the FR case,
the problem with Baumann and Brukner's analysis is that they fail to treat the
friend as an agent on the same footing as Wigner. If, instead, Wigner and his
friend are treated symmetrically, the BB scenario loses its seemingly
paradoxical character. Because the BB scenario is much simpler than the FR
scenario, we will discuss it first. Indeed it was through thinking about the BB
thought experiment that we finally arrived at our present understanding of the
more intricate FR thought experiment.

Our paper is organized as follows. In Section~\ref{sec:QBism} we
summarize the main QBist principles. We spell out what we mean by a user of the
quantum formalism, and how quantum states and quantum
measurements are thought of as personal judgments in our framework. We further explain how Asher Peres's dictum
that unperformed experiments have no results remains true even when an agent is certain of what he will find. In Section~\ref{sec:W} we review Wigner's original thought experiment and explain what it means for one agent to be a physical system
for another agent, distinguishing our notion from Wigner's original where he argued that the friend must be described as in a ``suspended
animation'' unless the laws of physics are changed. Section~\ref{sec:BB} contains the main argument of the paper. It shows that the BB
thought experiment can be understood fully within the standard quantum formalism if Wigner and his friend are treated in a fully symmetric
fashion. We turn to the FR thought experiment in Section~\ref{sec:FR}. We show that, exactly as in the BB case, the apparent contradiction
derived by Frauchiger and Renner is due to a failure to treat one of the participants in the thought experiment as an agent in the full sense
of the word. Finally we clarify the circumstances in which one agent may adopt another agent's quantum state assignments and thereby address a
challenge posed by Frauchiger and Renner~\cite{FR} in their subsection titled ``Analysis within QBism.''

% FR: "Nevertheless, there should be ways for agents to consistently reason about each other.

% \section{Agents and QBism}
\section{Agents and QBism}  \label{sec:QBism}

As there exist several authoritative accounts of QBism \cite{FuchsStacey2018,Fuchs2017,Stacey2019a}, this
section focuses on those aspects of QBism that are important for our argument
about agents and Wigner's friend. We start by defining the terms ``agent'' and
``user of quantum mechanics'' and discuss some key tenets of QBism. We then
give an account of the elementary double-slit experiment in QBist terms, in order to set
the scene for the discussion of Wigner's friend in the next section.

%* \subsection{Agents and users of quantum mechanics}
\subsection{Agents and users of quantum mechanics}
\label{subsec:agents}

According to QBism, the quantum formalism is a tool decision-making agents can adopt to better guide their decisions when faced with the inevitable uncertainties of the quantum world. Particularly, the theory guides its users in how to gamble on the personal consequences of their measurement actions. Thus for QBism, the quantum formalism plays a {\it normative\/} role for its users, not a descriptive role for exactly how the world is:  It suggests how a user {\it should\/} gamble.

Users of the theory are thus at the center of the QBist approach. It is therefore important to spell out what we mean by the term. In
the following, we will make a distinction between agents and users of quantum
mechanics:
\begin{itemize}
\item
{\it Agents\/} are entities that can take actions freely on parts of the world external
to themselves, so that the consequences of their actions matter for them.
\item
A {\it user of quantum mechanics\/} is an agent that is capable of applying the
quantum formalism normatively.
\end{itemize}

While our definition of a user is narrow, our definition of an agent is broad:
It does not rule out attributing agency to dogs, euglenas, or artificial
life. However, it does exclude a computer program that deterministically
``chooses'' an action from a look-up table. On the other hand, as Khrennikov
emphasizes in Ref.~\cite{Khrennikov17}, ``The idea is that QM is something used
only by a privileged class of people. Those educated in the methods of QM are
able to make better decisions (because of certain basic features of nature)
than those not educated in the methods of QM.'' This notion of a user of the
theory is sufficiently open to allow for additional details in the
future, but it is also precise enough for the purposes of this paper.

There exists a range of definitions of agency in the philosophical literature that overlap with our definition to
different degrees~\cite{Mueller2018,Briegel2012}.  According to the above definitions, a team of scientists sharing notebooks,
calculations, observations, etc., can act as a single agent and even a user of quantum mechanics~\cite{DeBrotaStacey2018}.

%% "Group of physicists agreeing on state: if they don't, they can't act as a single user if qm"

%% KEY IDEA TO BE MADE PROMINENT: gamble on the outside world, clear separation of
%% agent and system, no gambling if you can influence the outcome.

%% IMPORTANT NOTE: We will see that this does not bar the friend from using quantum
%% mechanics. The friend just has to do it right. The apparent paradox unveiled by
%% FR and BB shows that ``their'' friends are doing it wrong.

%% QM is not descriptive, does not tell a system what to do

%* \subsection{Some tenets of QBism}
\subsection{Some tenets of QBism}

The previous subsection started with a one-paragraph summary of QBism. The following
five tenets provide more detail. Taken together they ensure QBism's consistency.

%** \subsubsection{What is a measurement?}
\subsubsection{What is a measurement?} \label{sec:measurement}

A measurement is an action of an agent on its external world,
where the consequences of the action, or its {\it outcomes}, matter to the agent.

Like our definition of agent before, this definition of measurement is very
broad. Basically anything an agent can do to its external world---from opening
a box of cookies, to crossing a street, to performing a sophisticated quantum
optics experiment---counts as a measurement in our sense. The only thing that
sets quantum measurement as normally construed apart from the more mundane examples
given above is whether it is fruitful or worth one's while to apply the quantum
formalism to guide one's actions.  But, in principle a user of quantum
mechanics could use the formalism to make decisions in any measurement
situation, including measurements on living systems as in the Wigner's friend
thought experiment that are of concern here.

By applying the term measurement only to actions on the agent's external world, we exclude the case where an agent, directly or indirectly, acts on him or herself. We thus require a strict separation between the agent performing the measurement and the measured system.

%** \subsubsection{Measurement outcomes are personal}
\subsubsection{Measurement outcomes are personal}

When an agent performs a measurement---that is, takes an action on its external world---the ``outcome'' of the measurement is the consequence of
this action for the agent. A measurement outcome is personal to the agent
doing the measurement.  Thus two agents cannot experience the same
outcome. Different agents may inform each other of their outcomes and thus
agree upon the consequences of a measurement, but a measurement outcome should
not be viewed as an agent-independent fact which is
available for anyone to see \cite{Caves07}.

This tenet has led some commentators to claim mistakenly that QBism is a form
of solipsism. This claim has been thoroughly refuted (see, e.g.,
Ref.~\cite{Fuchs10a,Timpson08a,Mermin2014,Fuchs16,Healey2016}). That QBism is not solipsism follows immediately from the
premise that a measurement is an action on the {\it world external to the agent}. A
QBist assumes the existence of an external world from the outset.  Furthermore,
the consequences of measurement actions are beyond the agent's control---the
world can surprise the agent. The world is thus capable of genuine novelty complementary to the agent's actions---the world and the agent cannot be identified with each other. (See Refs.\ \cite[pp.\ 6--10]{Fuchs16} and \cite[pp.\ 19--20]{Fuchs2017}, arXiv versions.)

%** \subsubsection{A quantum state is an agent's personal judgment}
\subsubsection{A quantum state is an agent's personal judgment}

In QBism, the only purpose of the quantum formalism is to help an agent make better decisions. Rigorous use of the formalism enables an agent to make more successful gambles. The term ``gamble'' evokes games of luck, but here it is meant to encompass any action by an agent where the consequences matter to the agent. Any physics experiment is thus a gamble in this sense.

As we will explain in more detail in the next subsection, the quantum formalism can be viewed as an addition to classical decision
theory~\cite{Fuchs13a,Fuchs2017}. Following the approach to decision theory pioneered by Savage
\cite{Savage1972,BernardoSmith1994}, QBism takes all probabilities, including those equal to zero and
one, to be an agent's personal degrees of belief concerning future measurement outcomes.
Personalist probabilities \cite{deFinetti1990,Berkovitz2019} acquire an operational meaning by their use in decision making. A key consequence of this theory is that, to avoid sure loss, an agent's gambles must be constrained by the rules of probability theory.

In QBism, a quantum state is also an agent's personal judgment, reflecting the
agent's degrees of belief in the outcomes of all possible measurements he or she
might perform.  A quantum state, rather than being a property of a quantum
system, thus encodes an agent's expectations regarding the outcomes of future
measurements.

\subsubsection{The quantum formalism is normative rather than descriptive}
\label{sec:normative}

We will see below that the quantum-mechanical Born rule can be viewed as
placing additional constraints on an agent's probability assignments to
the outcomes of different measurements~\cite{Caves07} in situations where pure probability theory is simply silent.
In line with the central place that QBism gives to measurement, QBism treats
the Born rule as fundamental: To understand the quantum formalism, one has
to understand the Born rule first.

For a measurement with outcomes labeled $j=1,\ldots,n$, the Born rule is
usually given in the form $p_j={\rm tr}(\rho E_j)$, where $E_j$ is a
measurement operator or {\it effect\/} corresponding to outcome $j$ (for a von
Neumann measurement this will be a projection operator), $\rho$ is the density
operator for the measured system, and $p_j$ is the probability for outcome
$j$. In QBism, $\rho$ represents the agent's belief about the system, and the
list of effects (or POVM) $\{E_1,\ldots,E_n\}$ represents the agent's belief
about the measurement.

In contrast to the usual reading of the Born rule as a
formula for computing $p_j$ given $\rho$ and $E_j$, in QBism the Born rule
functions as a consistency requirement~\cite{Fuchs10a,Fuchs13a}. If an agent has beliefs $p_j$, $\rho$
and $E_j$ that do not satisfy the Born rule, he or she should modify at least
one of these beliefs. The formalism does not prescribe which one to modify or
how to modify it.\footnote{It is easier to see how $p_j$, $\rho$, and $E_j$ are on an equal footing if
$\rho$ and $E_j$ are expressed as probabilities. That this can be done is well
known: with respect to an appropriately chosen informationally complete
measurement, any density operator is equivalent to a vector of probabilities~\cite{Caves02b},
and any POVM $\{E_1,\ldots,E_n\}$ is characterized by a stochastic matrix of
conditional probabilities~\cite{Fuchs13a}.}

In QBism the Born rule is thus a consistency criterion that an agent should strive to satisfy in its probability and quantum state assignments.  It is a single-agent criterion; it says nothing about consistency between the probability and quantum state assignments of different agents.  It is entirely about internal consistency of an agent's expectations. This is what is meant by saying that the Born rule, and thus the quantum formalism, is normative rather than descriptive.

This tenet has led some commentators to claim mistakenly that QBism is a form
of instrumentalism. This, as with the claim of solipsism, is also easily
refuted; see, e.g., Refs.~\cite{Fuchs16,Wallace2016}. Indeed from its earliest
days, the very goal of QBist research has been to distill a statement about the
character of the world from the fact that gambling agents should use the
quantum formalism~\cite{Fuchs2002}. Even though this remains an ongoing
project, it has already led to a number of
strong ontological claims on the part of QBism---from the world being capable
of genuine novelty and being in constant creation, to the Born rule expressing
a novel form of structural realism~\cite{Fuchs2017,Fuchs16}.

\subsubsection{Probability-1 assignments are judgments}
\label{subsec:probOne}

QBism regards even probability-1 (and probability-0) assignments as an agent's personal
judgments. Assigning probability-1 to an outcome expresses the agent's
supreme confidence that the outcome will occur, but does not imply that
anything in nature guarantees that the outcome {\it will\/} occur.

Similarly, QBism regards both pure and mixed quantum states as an agent's
personal judgments. This implies in particular that even a statement such as
``this outcome is certain to occur'' reflects an agent's judgment rather than
a fact of nature. In other words, nothing in nature guarantees that an outcome
to which an agent has assigned proobability-1 will in fact occur.

\subsection{Unperformed measurements have no outcomes, even when an agent is
  certain what the outcome will be}

In his 1964 {\sl Lectures on Physics}, Richard Feynman famously stated that the
double-slit experiment exhibits ``the basic peculiarities of all quantum
mechanics''. ``In reality, it contains the {\it only\/} mystery,'' he said.
This idea seems to overlook the importance of quantum-foundational results such
as Bell inequalities, Kochen-Specker style noncolorability theorems, or
contextuality inequalities. Yet, in the end QBism believes Feynman was on the
right track.  Not only does the double-slit experiment exhibit the basic
peculiarities of quantum mechanics, it points to the solution of the Wigner's
friend conundrum as well.

The double-slit experiment consists of a particle source, a screen with two
slits, and a second screen farther away from the source where the particle
position is recorded. Assume an agent has made assignments $P(H_0)$
and $P(H_1)$ for the probability that the particle passes through the left or
right slit, respectively, and $P(D_j|H_0)$ and $P(D_j|H_1)$ that the particle is
detected at position $D_j$ given that it passes through the left or right slit,
respectively. We assume $P(H_0)+P(H_1)=1$ and $\sum_j P(D_j|H_0)=\sum_j P(D_j|H_1)=1$, as must be the case for probabilities and conditional probabilities.
As explained above, for these probabilities to have operational,
decision-theoretic, meaning they have to refer to actual outcomes for the
agent. Spelled out, this means $P(H_0)$ is the probability that the agent sees
the particle pass through the left slit, allowing the agent to gamble on this
outcome, and $P(D_j|H_0)$ is the probability that the agent detects the particle at $D_j$
given that he or she has seen it pass through the left slit, allowing the agent
to make the corresponding conditional gamble. (The same applies, of course, to
the right slit.)

In the uncontroversial case where the agent actually intends to check which
slit the particle passes through before it hits the second screen, the agent's
probability $P(D_j)$ for finding the particle at $D_j$ is given by
\be
P(D_j) = P(H_0) P(D_j|H_0) + P(H_1) P(D_j|H_1)\;. \label{eq:totprob}
\ee
This follows from probability theory alone.

But what if the agent does not intend to check which slit the particle passes
through? In this case we are dealing with a {\it different experiment\/} for
which probability theory alone does not constrain the agent's
probabilities. The classical intuition in this situation is to continue to use
Eq.~(\ref{eq:totprob}) for the probability of detecting the particle at $D_j$,
which amounts to the physical postulate that as far as this probability is
concerned, it does not matter whether the agent does or does not check which
slit the particle goes through.

This classical intuition has to be abandoned in quantum mechanics. Whether a
measurement is performed or not matters profoundly. Asher Peres expressed this
in his famous slogan ``Unperformed experiments have no results.'' \cite{Peres78}

Crucially, in a quantum analysis of the double-slit
experiment, there is no change in either the values or the meaning of the
probabilities $P(H_0)$, $P(H_1)$, $P(D_j|H_0)$, and $P(D_j|H_1)$. For instance, $P(D_j|H_0)$ is
still the agent's probability to detect the particle at $D_j$ given that the
agent saw it pass through the left slit. In his paper ``The Concept of
Probability in Quantum Mechanics'' \cite{Feynman51}, Feynman makes a
similar point and then goes on to write that ``[w]hat is changed, and changed
radically, is the method of calculating probabilities.''

Here QBism departs from Feynman's view in one significant way. In the uncontroversial case
that the agent intends to do an intermediate measurement to check which slit
the particle goes through, the quantum rules do not lead to a change of the
probabilities $P(D_j)$ as given in Eq.\ (\ref{eq:totprob}). If the agent does not
intend to do the intermediate measurement, however, Eq.~(\ref{eq:totprob}) no
longer applies because the probabilities $P(H_0)$,
$P(H_1)$, $P(D_j|H_0)$, and $P(D_j|H_1)$ now refer to a {\it hypothetical\/} intermediate
measurement. Thus probability theory alone no longer
gives a formula for the probability of finding the particle at $D_j$. The
``radically'' new quantum method of calculating probabilities in this case
should therefore not be viewed as a change of, but as an addition to, existing
methods. Probability theory remains fully valid in the quantum
realm.

%% JOHN WRITES: Not as a change to probability theory, an empirically motivated
%% physical addition to it, just as the classical intuition was. But the
%% classical intuition potential addition is replaced, not added to. Careful to
%% note that we need not necessarily change expectations we had for various
%% things, even those that we adopt in light of the "wrong" physical
%% assumption. Many can be accommodated, etc.

As we stated in Section \ref{subsec:probOne}, QBism takes the stand that even
when an agent assigns probability-1 to one of the possible outcomes of a
measurement, there is nothing in the agent's external world that metaphysically
ensures it will necessarily come about. For ``unperformed measurements have no
outcomes'' is a statement about the assumed character of the world, whereas a
probability-1 assignment is only a belief (supremely strong, but nonetheless
a belief) that someone happens to have in the moment. That unperformed
measurements have no results is, for QBism, the great lesson of all the
Bell-inequality and Kochen-Specker-theorem results of the last half century,
more recently reinforced by the ``no-go theorems'' of Pusey, Barrett, and
Rudolph~\cite{PBR12} and Colbeck and Renner \cite{Colbeck2012}.
It plays a central role  in
the QBist approach to Wigner's friend.

\section{Wigner's Friend}    \label{sec:W}

Wigner described his thought experiment in a 1961 paper entitled ``Remarks on
the Mind-Body Question''~\cite{Wigner1961}. Below we use a slightly modernized version of
Wigner's notation. The friend (who prefers the pronouns ``she'' and ``her'')
performs a two-outcome measurement on a quantum system, where the outcomes
correspond to the states $|\psi_1\rangle$ and $|\psi_2\rangle$,
respectively. In order to be consistent with the BB scenario discussed in
Section~\ref{sec:BB} below, we assume that $|\psi_1\rangle$ and
$|\psi_2\rangle$ are states of a spin-1/2 particle corresponding to ``spin up''
and ``spin down'', respectively. After the friend's measurement, Wigner
contemplates a simple measurement on her, consisting of the question: what was
the result of your spin measurement?

The assumption is now that Wigner assigns a quantum state to the joint
system consisting of particle and his friend, and treats it as a closed
quantum system. After the friend has measured the spin, Wigner's joint state becomes
\be
|\Phi\rangle = \alpha |\psi_1\rangle |\chi_1\rangle + \beta
|\psi_2\rangle |\chi_2\rangle \;, \label{eq:phi}
\ee
where $  |\chi_1\rangle$ and $|\chi_2\rangle$ are Wigner's states for his friend and correspond to her responding
to the question ``what was the result of your spin measurement'' with ``up''
and ``down'', respectively.

In his 1961 paper, Wigner argued that the question of whether the friend saw
``up'' or ``down'' was already decided in her mind, before Wigner asked her.
Here is what Wigner concludes from this, in his own words (excepting our
modernized notation):

\bq
If we accept this, we are driven to the conclusion that the proper wave
function immediately after the interaction of friend and object was already
either $|\psi_1\rangle |\chi_1\rangle$ or $|\psi_2\rangle |\chi_2\rangle$ and not the linear
combination $\alpha |\psi_1\rangle |\chi_1\rangle + \beta |\psi_2\rangle |\chi_2\rangle$.  This
is a contradiction, because the state described by the wave function
$\alpha |\psi_1\rangle |\chi_1\rangle + \beta |\psi_2\rangle |\chi_2\rangle$
describes a state that has
properties which neither
$|\psi_1\rangle |\chi_1\rangle$ nor $|\psi_2\rangle |\chi_2\rangle$
has. If we substitute for ``friend'' some simple physical apparatus, such as an atom
[\ldots], this difference has observable effects and {\it there is no doubt
that $\alpha |\psi_1\rangle |\chi_1\rangle + \beta |\psi_2\rangle |\chi_2\rangle$
describes the properties of the joint system correctly, the assumption that the wave
function is either $|\psi_1\rangle |\chi_1\rangle$ or $|\psi_2\rangle |\chi_2\rangle$ does
not} [Wigner's italics]. If the atom is replaced by a conscious being, the
wave function $\alpha |\psi_1\rangle |\chi_1\rangle + \beta |\psi_2\rangle |\chi_2\rangle$ (which
also follows from the linearity of the equations) appears absurd because it
implies that my friend was in a state of suspended animation before [she] answered
my question.

It follows that the being with a consciousness must have a different role in quantum mechanics than
the inanimate measuring device: the atom considered above.
In particular, the quantum mechanical equations of motion cannot be
linear if the preceding argument is accepted.  This argument implies that ``my
friend'' has the same types of impressions and sensations as I---in particular,
that, after interacting with the object, [she] is not in that state of suspended
animation which corresponds to the wave function $\alpha |\psi_1\rangle |\chi_1\rangle + \beta |\psi_2\rangle |\chi_2\rangle$.
\eq

QBism, along with most other current interpretations of quantum mechanics, does
not restrict the applicability of quantum mechanics to inanimate devices. An
agent can apply the normative quantum calculus to any part of the world
external to him or herself, including conscious beings and other agents. So how
does QBism escape the conclusions that Wigner draws from his thought
experiment?

Part of the answer is straightforward and follows directly from the QBist
tenets. Wigner's quantum state $|\Phi\rangle$ is not descriptive: it does not
describe properties of the joint system to which it refers. The exclusive role
of $|\Phi\rangle$ is to help Wigner quantify his expectations regarding the
consequences of his future actions on friend and particle. In the same way,
the friend's state assignments refer to her expectations regarding
the consequences of her future actions. There is simply no conflict between
these two perspectives. In particular, Wigner's state assignment has no
bearing on whether the friend is or is not in a ``state of suspended
animation''.

This straightforward resolution of Wigner's paradox has profound implications
for the QBist worldview. A few lines below the quoted passage, Wigner points
out that insisting on the superposition state $|\Phi\rangle$ for particle and
friend, though not necessarily a contradiction, amounts to denying ``the
existence of the consciousness of a friend'' to an intolerable extent. In
QBism, a quantum state assignment has no bearing on the existence of the
consciousness of a friend. It follows that a QBist can simultaneously assign
the state $|\Phi\rangle$ and grant his friend a conscious experience of having
seen either ``up'' or ``down''.

This claim requires some elaboration. The scenario of Wigner's friend can be
understood as a version of the double-slit experiment, in line with
Feynman's dictum that the latter contains the basic peculiarities of all
quantum mechanics. As in our analysis of the double-slit experiment, none of
the probabilities considered in Wigner's paper change their value or their meaning
when Wigner writes down his quantum state $|\Phi\rangle$. The only implications
of Wigner assigning a quantum state to the friend are that, (i) as far his probabilities for the outcomes
of some future quantum measurement on the friend are concerned, it matters
whether or not he first asks her whether she saw up or down, and (ii) that he should use
the Born rule to compute these probabilities. There is no reason why Wigner
cannot assign a quantum state that respects all of his beliefs about the
friend's inner life, conscious experiences, or agenthood.

The parallel with the double-slit experiment is somewhat hidden in Wigner's
original argument, because he only considers measurements on the friend that,
in the double-slit experiment, correspond to determining which slit the
particle went through. But, as Wigner makes clear when he writes that ``there is
no doubt that [$|\Phi\rangle$] describes the properties of the joint system
correctly'', assigning the state $|\Phi\rangle$ amounts to committing to
predictions for the outcomes of a wide range of quantum measurements on the
friend, including those for which the predictions depend on whether or not the
friend is first asked what she saw. In the context of the Wigner's friend
scenario, such measurements were first considered by David Deutsch
\cite{Deutsch1985}. They are crucial for the BB thought experiment,  to which
we will now turn.

\section{The Friend's Perspective: Response to Baumann \& Brukner}  \label{sec:BB}

%% Wigner's original is simply the double slit experiment. Unperformed experiments
%% have no results.
%%
%% Frauchiger and Renner: Rejection of C in general follows from "unperformed
%% experiments have no outcomes even for p=1". When can you use informal
%% reasoning, classical intuition? When you judge that looking doesn't matter. But
%% the thought experiment brings out a more interesting aspect of QBism:
%%
%% Can two agents see the same outcome? (not two experimenters who form part of a
%% team). No and yes. No in the sense of "friend's outcome is not `Wigner saw
%% w=OK'".  Yes in the sense of Copernican principle, the you.
%%
%% The fault: where F predicts w=OK.
%%
%% 1. BB innovation: friend makes prediction for Wigner's outcome
%%
%% 2. FR innovation: identify clear principle/assumption needed for Wigner to take
%%    into account friend's prediction (but then it's about the same outcome)
%%
%% 3. Both BB and FR assume that there is one correct quantum state for the setup,
%%    which happens to coincide with Wigner's perspective
%%
%% 4. In both BB and FR, friend assigns state to herself, which is not correct
%%    whenever state is viewed as relative to observer
%%
%% 5. In both BB and FR, it matters that two agents cannot see the same thing.
%%
%% 6. (a) symmetry (b) what it means and does not mean to assign a quantum state
%%    to an agent (c) what is an agent in the theory (d) Renato's challenge: how
%%    and when does one agent take another agent's view into account?
%%

Baumann and Brukner's thought experiment~\cite{BB} is a simple modification of Wigner's
original scenario. After the friend has made her measurement, Wigner's joint
state of particle and friend is again given by Eq.\ (\ref{eq:phi}), where
it is now assumed that $\alpha=\beta=1/\sqrt2$, so that we have
\begin{equation}
|\Phi\rangle = \frac1{\sqrt2}\Big(|\psi_1\rangle |\chi_1\rangle +
|\psi_2\rangle |\chi_2\rangle \Big) \;.
\end{equation}
Whereas in Wigner's paper, Wigner contemplates a simple measurement on the friend
consisting of asking her about the result of her spin measurement, Baumann and
Brukner let Wigner do the measurement
\begin{equation}
  M_W:\;\{|\Phi\rangle\langle\Phi|, 1-|\Phi\rangle\langle\Phi|\} \;.
\end{equation}
Such a measurement is far beyond any current and probably future experimental
possibilities, but if we allow Wigner to write down the state $|\Phi\rangle$, we must
also allow him to contemplate the measurement $M_W$. Clearly, Wigner has a
probability $p=1$ of obtaining the outcome corresponding to
$|\Phi\rangle\langle\Phi|$, which is labeled ``+'' in Ref.~\cite{BB}.

Baumann and Brukner's main claim concerns the friend's prediction for the ``+''
outcome. They argue that, if in her spin measurement the friend obtains ``up'',
her probability of ``+'' is given by applying $M_W$ to the state
$|\psi_1\rangle |\chi_1\rangle$, and if she obtains ``down'',
she should apply $M_W$ to the state
$|\psi_2\rangle |\chi_2\rangle$. In both cases, her probability for ``+''
is 1/2. Since this probability is the same for ``up'' and ``down'', she can
communicate her prediction to Wigner without affecting the rest of the
experiment. Baumann and Brukner's claim thus leads to the troubling conclusion
that two different ways of applying the rules of quantum mechanics
give contradictory numbers for the probability of ``+''. \footnote{Baumann and Brukner
assume the experiment is repeated many times, so that the alleged contradiction can be phrased in frequentist
terms. From a QBist perspective, the full force of the contradiction arises
already in the single-case analysis given here.}

In the above account it might appear problematic that, in QBism, the outcome of the measurement
$M_W$ is personal to Wigner. But Baumann and Brukner show a valid way around
this problem by stipulating that Wigner record the outcome of his measurement
on a piece of paper.  The friend's probability assignment can then be regarded
as referring to her finding ``+'' upon checking the piece of paper.

The real problem with the BB analysis is that for the friend to base her
prediction on the state $|\psi_1\rangle |\chi_1\rangle$ (or $|\psi_2\rangle
|\chi_2\rangle$) amounts to assigning a quantum state to herself, which
violates the QBist tenet
that there must be a clear separation between agent and measured system.  It is
easy to see why this leads to a serious difficulty. Assume for the moment that
the friend writes down $|\psi_1\rangle |\chi_1\rangle$ for the joint
system of particle and herself and uses it to compute her probability for what
she will see on the piece of paper. This state assignment would not just commit
her to a probability for outcome ``+''. It would commit her to probabilities
for any measurement that she could perform on the particle and herself. For
instance, according to our discussion in Section~\ref{sec:W}, the state
$|\chi_1\rangle$ corresponds to her responding with ``up'' to the question
``what was the result of your spin measurement''. But since she is a free
agent, she has control over the answer to this question. It is up to her
whether she replies ``up'', ``down'', or by sticking her tongue out. Since she
has at least partial control over these measurement outcomes, the above
quantum-state assignment cannot form a reliable basis for guiding her actions.

So what should the friend do instead? The answer has already been given in the
Introduction and Figure 1. Rather than adopting Wigner's viewpoint, she needs
to analyze the experiment as an action that she takes on the particle, the lab,
Wigner, and the piece of paper on which Wigner records his outcome. In
particular, this requires her reversing roles and treating Wigner as a physical
system. That this would be an enormously complex and practically infeasible
task is hardly a valid objection given the assumption that Wigner is able to
write down a quantum state for her. Indeed, an even-handed analysis of the
thought experiment clearly requires the assumption that the friend is as
skillful a user of quantum mechanics as Wigner himself.

What probability should the friend assign to her finding ``+''
on the piece of paper?  This depends on her prior states, unitaries, and POVMs
regarding the lab and Wigner. The only constraint on her probability for ``+''
is that it should be consistent with her prior quantum assignments in the sense
given in Section~\ref{sec:normative}, i.e., it should be consistent with the
Born rule.  This implies that the friend's probability for ``+'' cannot be
derived from the details provided in the BB thought experiment. Furthermore, if
the experiment is repeated many times as envisaged by Baumann and Brukner, the
formalism will typically lead her to update her assignments after each
repetition. Her probabilities will thus reflect what she learned in previous
runs of the experiment.

Here is a summary of our argument. In the same way that Wigner does not take
the friend's viewpoint into account when he computes his probability of ``+'',
the friend need not take Wigner's viewpoint into account when she computes her
probability of ``+''. This puts Wigner and the friend on an equal footing.  In
particular, the friend's quantum state assignments is not a function
of Wigner's quantum states $|\chi_1\rangle$ and $|\chi_2\rangle$. We thus
explicitly reject Baumann and Brukner's claim that standard quantum theory
requires the friend to base her probability assignments on $|\chi_1\rangle$ and
$|\chi_2\rangle$. In contrast to Baumann and Brukner, who propose that the
friend uses a modified Born rule incorporating Wigner's perspective, our
symmetric QBist treatment of Wigner and his friend requires no such
modification.

%% The lesson to be learnt from Baumann and Brukner's thought experiment is that
%% the friend should not adopt Wigner's state assignment to herself.  This gives a
%% partial answer to the question asked in the introduction: Under what
%% circumstances, and in what way, should one agent adopt another agent's quantum
%% state assignments? In the next section, we will address the this question from
%% Wigner's perspective.

%% [JOHN writes: Do we want to also point out in this section that the friend
%% could, entirely without the use of quantum theory in the particular case of
%% being in the experiment, merely have the same probabilities as Wigner due to it
%% being in her prior beliefs? The point being that this is not a use of quantum
%% theory, but just a reflection of expectations she has for what happens when
%% people go into the box.]

% \section{Reasoning about other agents: The FR thought experiment}
\section{Reasoning about Other Agents: Response to Frauchiger \& Renner} \label{sec:FR}

Frauchiger and Renner's thought experiment, which considers four agents, is
somewhat intricate, but here only the following broad outline is
needed. Two agents, $\Friendtwo$ (the friend) and $\Friendone$, are located in
separate labs. The other two agents, $\Wigner$ (Wigner) and $\Assistant$, are
on the outside and perform measurements on the labs. At time $t=0$, agent
$\Friendone$ prepares a qubit in a given state, measures it, prepares a
spin-1/2 particle in a state that depends on the measurement outcome, and sends
the particle to agent $\Friendtwo$'s (the friend's) lab. At $t=1$, the friend
measures the particle. At $t=2$, agent $\Assistant$ measures
$\Friendone$'s lab in a given basis. Finally, at $t=3$, Wigner measures the
friend's lab in a given basis.

Wigner now uses two different methods to make predictions for the outcome of
his measurement. He is interested in the probability of one of the outcomes,
labeled $w$=fail. His first prediction uses the quantum formalism. For his
second prediction, he reasons about what predictions the other agents would
have made at earlier stages of the experiment, assuming all agents start from
the same initial quantum state assignment. Frauchiger and Renner argue that
the two methods lead to mutually contradictory predictions.

When reasoning about other agents, Wigner, agent $\Assistant$, and the friend
apply FR's
\bq
\noindent{\bf Assumption (C)}: Suppose
        that agent A has established that ``I am certain that agent A$'$, upon
        reasoning within the same theory as the one I am using, is certain that
        $x=\xi$ at time $t$.'' Then agent A can conclude that ``I am certain that
        $x=\xi$ at time $t$.''
\eq
The FR argument starts with agent $\Friendone$ making, immediately after time
$t=0$, a prediction about Wigner's measurement outcome. Since the measured
system is the friend's lab, agent $\Friendone$'s prediction is about a part of
her external world. The next step is that the friend applies Assumption (C) to
make $\Friendone$'s prediction her own. Then $\Assistant$ applies Assumption
(C) to make the friend's and thus $\Friendone$'s prediction his own. Finally
Wigner applies Assumption (C) to make $\Assistant$'s and thus also the friend's
and $\Friendone$'s prediction his own. These steps are cleverly arranged
in time so that they don't clash with the different measurements.

Notice that all four agents' predictions concern the same outcome, namely
$w$=fail in Wigner's measurement of the lab containing the friend. This means
in particular that the FR argument depends on the friend making a prediction
about Wigner's measurement on herself. In their table 3, Frauchiger and Renner
make this explicit by stating the friend's conclusion as ``I am certain that
[Wigner] will observe $w$=fail at time [$t=3$].''

But this means we are now in the same situation as in the previous section when
we analyzed the BB thought experiment. The friend {\it can\/} use the quantum
formalism to make a prediction for Wigner's outcome (more precisely, for what
she will find when she checks a record of Wigner's outcome). But she is not
required to base her prediction on agent $\Friendone$'s or Wigner's state
assignments. She will have to analyze the experiment as an action that she
takes on the other lab and the other agents. Similar to the discussion of the
BB thought experiment, the contradiction derived by FR is resolved if the
symmetry between Wigner and his friend is recognized.

Frauchiger and Renner state correctly that Assumption (C) is rejected by
QBism. This does not mean that there is a prohibition in QBism for one agent to
adopt another agent's probability or state assignments.  A QBist agent will
have to decide on a case by case basis whether or not to do so. A
straightforward way of making use of other agents' probabilities in one's
decision making is simply to ask them what their probabilities are and to treat
the answers as data which they may or may not take into account in their own probabilities.

The most common scenario in which scientists adopt each others' probability and
state assignments is that of a team working jointly on a quantum experiment and
acting as a single agent and user of quantum mechanics. It follows from the
definitions given in Section \ref{subsec:agents} that scientists in such a team
must have common probability and state assignments. The requirement of a strict
separation between agent and measured system now translates into a strict
separation of team and measured system. For the FR thought experiment this
means that its four players cannot be thought of as acting as a single agent,
because they perform measurements on each other.

In a subsection of their paper, titled ``Analysis within QBism'', Frauchiger and
Renner write that ``Nevertheless, there should be ways for agents to
consistently reason about each other.'' In this paper we have provided such a
way. For two users of quantum mechanics who interact, it requires each of them
to treat their interaction as an action he or she takes freely on the other.

% \section{Conclusion}
\section{Conclusion}

We have seen that the thought experiments described by Frauchiger and Renner
and by Baumann and Brukner have a key aspect in common. In each of them, an
agent (the friend) is using quantum mechanics to predict the consequences of an
action performed on her by another agent (Wigner). We have shown that the
paradoxa found by these authors disappear if the friend analyses the experiment
as an action she performs on the world outside herself, which includes Wigner.
These thought experiments thus illustrate what we have called a quantum
Copernican principle: when two agents take actions on each other, each agent
has a dual role as a physical system for the other agent.

% \section*{Acknowledgements}
\section*{Acknowledgements}

We would like to thank Renato Renner, \v{C}aslav Brukner, Veronika Baumann, and
Jacques Pienaar for many valuable discussions.  CAF was supported in part by the John E. Fetzer Memorial Trust; CAF and JBD were further supported by grant FQXi-RFP-1811B of the Foundational Questions Institute.

\end{document}